\documentclass{nature}

\def \ga{\mathrel{\mathchoice {\vcenter{\offinterlineskip\halign{\hfil$\displaystyle##$\hfil\cr>\cr\sim\cr}}}
{\vcenter{\offinterlineskip\halign{\hfil$\textstyle##$\hfil\cr>\cr\sim\cr}}}
{\vcenter{\offinterlineskip\halign{\hfil$\scriptstyle##$\hfil\cr>\cr\sim\cr}}}
{e\vcenter{\offinterlineskip\halign{\hfil$\scriptscriptstyle##$\hfil\cr>\cr\sim\cr}}}}}

\title{A gravitationally lensed water maser in the early Universe}

\author{C. M. Violette Impellizzeri$^{1}$, John P. McKean$^{1}$, Paola Castangia$^{1,2}$, Alan L. Roy$^{1}$, Christian Henkel$^{1}$, Andreas Brunthaler$^{1}$ \& Olaf Wucknitz$^{3}$}
\begin{document}

\maketitle

\begin{affiliations}
\item Max-Planck-Institut f\"{u}r Radioastronomie, Auf dem H\"{u}gel 69, D-53121 Bonn, Germany
\item INAF-Osservatorio Astronomico di Cagliari, Loc. Poggio dei Pini, Strada 54, I-09012 Capoterra (CA), Italy
\item Argelander-Institut f\"{u}r Astronomie, Auf dem H\"{u}gel 71, D-53121 Bonn, Germany
\end{affiliations}

\begin{abstract}
Water masers$^{1-4}$ are found in dense molecular clouds closely
associated with supermassive black holes in the centres of active
galaxies.  Based upon the understanding of the local water maser
luminosity function$^{5}$, it was expected that masers at intermediate
and high redshifts would be extremely rare, but galaxies at redshifts
$z>2$ might be quite different from those found locally, not least
because of more frequent mergers and interaction events. Using
gravitational lensing as a tool to enable us to search higher
redshifts than would otherwise be possible, we have embarked on a
survey of lensed galaxies, looking for masers.  Here we report the
discovery of a water maser at redshift 2.64 in the dust- and gas-rich
gravitationally lensed type~1 quasar MG~J0414+0534$^{6-13}$, which, with an isotropic luminosity of 10\,000\,$L_\odot$, is twice as luminous as the most powerful local water maser$^{14}$, and half that of the most distant maser previously known$^{15}$. Using the 
locally-determined luminosity function$^{5}$, the probability of
finding a maser this luminous associated with any single active galaxy is $10^{-6}$. The fact that we saw such a maser in the
first galaxy we observed must mean that the volume densities and
luminosities of masers are higher at that epoch.
\end{abstract}

Observations of the H$_2$O $6_{16}-5_{23}$ transition (rest frequency
22.23508~GHz) from MG J0414+0534 (see Supplementary Information) were
made with the 100~m Effelsberg radio telescope at the redshifted
frequency of 6.1~GHz during July and September 2007. The radio
spectrum (Fig.~1, upper panel) shows an emission feature detected with
a signal-to-noise ratio of seven, which we identify as a water
maser. The emission arises from the amplification of background
photons by stimulated emission of water molecules that have been
pumped up to a long-lived excited state by collisional
excitation$^{4}$ (see Supplementary Information).  The line emission
cannot be associated with the lensing galaxy at redshift 0.958 or with
a nearby ``local'' object because there are no known strong lines at
the corresponding rest frequencies (11.975~GHz and 6.116~GHz,
respectively). 

To confirm the detection of water maser emission and to
match it spatially with the lensed quasar, interferometric
observations with the Expanded Very Large Array (EVLA) were made
during September and October 2007. A clear emission line was detected
with a signal-to-noise ratio of six (Fig. 1, bottom panel), when
integrated over the position of the two strongest images (A1+A2) of
the lensed quasar (Fig. 2). The maser line is also detected in the
separate spectra of A1 and A2, although to a lower signal-to-noise
ratio. The EVLA data were not sensitive enough to detect the water
maser emission from the two weaker lensed images (B and C). The radial
velocities measured with Effelsberg and the EVLA are identical to
within the uncertainties (see Supplementary Table 1). The water maser line is also
coincident in velocity with the blue-shifted peak of CO emission and
with the strongest H\,{\sc i} absorption trough previously reported
from MG~J0414+0534$^{12,13}$ (Fig.~1).  There is no evidence of water
maser emission at the velocities of the other known CO emission and
H\,{\sc i} absorption components. Our detection is consistent with a
previously reported non-detection that did not reach sufficient
sensitivity$^{16}$.

Hitherto, the highest redshift at which water had been observed was 0.66$^{15}$. The H$_2$O maser in MG~J0414+0534 is at a redshift of 2.639, which is a factor of six more distant (using luminosity distances).  The measured maser transition requires gas temperatures in excess of 300\,K and particle densities $n$(H$_2$) $\ga$ 10$^7$~cm$^{-3}$ (see Supplementary Information). The most dense gas previously observed at high redshift was HCO$^{+}$ in the Cloverleaf gravitational lens system$^{17}$, tracing a density of 10$^5$~cm$^{-3}$ in star-forming molecular clouds. The apparent unlensed luminosity, assuming
that the maser originates from the same region as the continuum core
emission and hence has approximately the same magnification
($\sim35$$^{9}$; see Supplementary Information), is of order
10\,000\,$L_\odot$. This luminosity is still extraordinarily high,
the most luminous water maser known being in a type 2 quasar at
redshift 0.66 with a luminosity of 23\,000\,$L_\odot$$^{15}$. In the
event that the water maser is not coincident with the radio continuum,
but lies closer (farther) to the lens caustic, the magnification could
be higher (lower). We note, however, that all luminous (relatively
nearby) water masers studied so far in detail are either associated
with the circumnuclear accretion disk or a relativistic jet in its
immediate vicinity.  

Even without lensing, the water maser in MG~J0414+0534 is among those
with the highest known apparent luminosities$^{14,15}$. Nevertheless,
this discovery was possible only due to the additional amplification
provided by the foreground galaxy; it acts as a cosmic telescope
reducing the integration time required for the detection by a factor
of order 1000. The probability of a single pointed observation like
ours detecting a high redshift water maser with a luminosity greater
than 10\,000 $L_\odot$ is only $10^{-6}$ (see Supplementary
Information). This detection probability was determined by
extrapolating the local water maser luminosity function$^{5}$ to
higher luminosities without a cut-off, and assuming that the
luminosity function at high redshift was the same as that
locally. Thus, our detection of a water maser in MG J0414+0534 at
redshift 2.64 rules out with a high confidence no evolution in the
water maser luminosity function, and requires that the space density
of luminous water masers was much larger at high redshift than in the local
Universe. However, systematic searches for this population of unmagnified water
masers at high-redshift will probably require a significant
improvement in instrument sensitivity.  The proposed Square Kilometer
Array (SKA), expected to be operational within a decade, will provide
such a dramatic improvement$^{1}$.

These apparent lensed and unlensed luminosities have been calculated
based on the standard assumption of isotropic emission of radiation,
for comparisons with the luminosity estimates of other
masers. However, masers are likely to emit
anisotropically$^{18}$. Beaming is expected since differences in the
gain path due to irregularities in the cloud shape and velocity
coherence cause exponential changes in maser output brightness.
Beaming may also occur in some cases due to the alignment of masing
clouds occurring only in restricted directions or due to competitive
pumping in saturated masers, in which the stimulated emission in one
directional mode dominates over other directional modes. The resulting
beamwidths are uncertain, but arguments have been made for values from
7 degrees$^{19}$ down to as low as milliarcseconds$^{20}$. This
expectation is testable for the first time using the gravitationally
lensed water maser since the light seen through each of the lensed
images was emitted in slightly different directions from the
background quasar. The angle subtended between image regions A1 and
A2, as seen from the quasar, is 0.5~arcseconds, and between images A1
and B is 2.3~arcseconds. The maser line is seen in the spectra of both
lensed images A1 and A2 from the EVLA observations and so the maser
beaming angle is greater than 0.5~arcseconds, ruling out
milli-arcsecond beaming angles for this system. Furthermore, the
intrinsic luminosity of the water maser must be greater than 5
$\times$ 10$^{-9}$ $L_\odot$.

Of the $\sim$100 galaxies known to host 22.2 GHz water masers$^{21}$,
most are in type~2 Seyfert or LINER (Low-Ionization Nuclear
Emission-Line Region) galaxies at redshifts $<$0.06$^{2-4,22}$. The
notable exception is a type~2 quasar at redshift 0.66$^{15}$.  This is
consistent with unification models$^{23}$ in which the type~2 optical
spectrum is due to an edge-on orientation of the circumnuclear disk,
causing the active nucleus to be hidden behind a large column of dust
and gas (see Supplmentary Information). This geometry provides a long
maser gain path-length for amplification and so the prevalence of
masers in type~2 active galactic nuclei fits naturally with
unification models. MG~J0414+0534 is an intriguing object as it is one
of the few type~1 active galactic nuclei and the only known type~1
quasar to show water maser activity.

At low redshifts, at least one third of known water masers are
associated with the orbiting molecular clouds of circumnuclear
accretion disks$^{21}$. These water masers are typically found within
0.1 to 1.0~parsec of the supermassive black hole and tend to have
multiple blue-shifted, red-shifted, and systemic velocity components,
where individual components have narrow line widths of
$<$5~km\,s$^{-1}$~$^{19,24,25}$. In three cases, luminous water masers
have also been found to be associated with the nuclear parts of relativistic jets that are ejected from some central engines$^{26-28}$. These masers have
relatively broad line widths, up to 100~km\,s$^{-1}$, and have
velocities that tend to be offset from the systemic velocity of the
host galaxy. Whether the water maser in MG~J0414+0534 is associated
with the circumnuclear accretion disk or is induced by a relativistic
jet interacting with a gas cloud is not conclusive from our data
alone. However, given that only a single emission line has been
detected, and that it is broad and offset from the systemic velocity
by $-300$~km\,s$^{-1}$, the jet-maser scenario seems most likely. The
type~1 optical spectrum and beamed radio continuum emission of the
quasar provide further support, since unification models of type~1
objects have the nucleus being viewed from above the plane of the
disk. Masers originating from the circumnuclear accretion disk,
however, are preferentially beamed in the plane of the
disk. Therefore, disk masers are unlikely to be seen in active
galactic nuclei of type~1, while masers associated with nuclear jets
may well be detectable.

Future high-resolution imaging of the water maser line with very long baseline interferometry (VLBI) will
provide the exact location of the emission relative to the core-jet
radio structure already observed in MG~J0414+0534$^{9,29}$.  Resolving
such maser component distributions with VLBI would usually be
challenging at cosmological redshifts since the angular resolution of
global VLBI arrays operating at 6.1~GHz is $\sim$2~milliarcseconds,
which corresponds to a spatial resolution of $\sim$15~parsec at
redshift 2.639. However, for MG~J0414+0534, the apparent angular
extent of the radio structure is increased due to the foreground
gravitational lens by a factor of $\sim$15 for the two strongest
lensed images, A1 and A2$^{9}$ (see Supplementary Information). Hence,
the spatial resolution of a VLBI image will be $\sim1$~parsec.
Furthermore, it will be possible to distinguish water masers separated
by $\sim$ 0.5~parsec in the background source if each masing component
is detected with a signal-to-noise ratio of at least ten, thus
matching or even resolving the $\ga$0.5~parsec outer diameter of known
maser disks in the Circinus galaxy and NGC~1068$^{28,30}$. Thus, VLBI
imaging of gravitationally magnified water masers has the powerful
potential to trace the sub-parsec scale structure surrounding
accretion disks at cosmological distances.

\begin{addendum}
\item Our results are based on observations with the 100-m telescope of the
MPIfR (Max-Planck-Institut f{\"u}r Radioastronomie) at Effelsberg and
the EVLA which is operated by the National Radio Astronomy Observatory
which is a facility of the National Science Foundation operated under
cooperative agreement by Associated Universities, Inc. The authors are
grateful to Alex Kraus and Mark Claussen who helped make these
observations successful. JPM was supported by the European Community's
Sixth Framework Marie Curie Research Training Network `ANGLES'. AB and
OW were supported by the Priority and Emmy-Noether-Programmes of the
Deutsche Forschungsgemeinschaft, respectively.

\item[Competing Interests] The authors declare that they have no competing financial interests.

\item[Correspondence] Correspondence and requests for materials
 should be addressed to C.M.V.I.~(email: violette@mpifr-bonn.mpg.de)
\end{addendum}

\newpage

\begin{figure}
\begin{center}
\setlength{\unitlength}{1cm}
\begin{picture}(6,10.2)
\put(-5,-6){\includegraphics{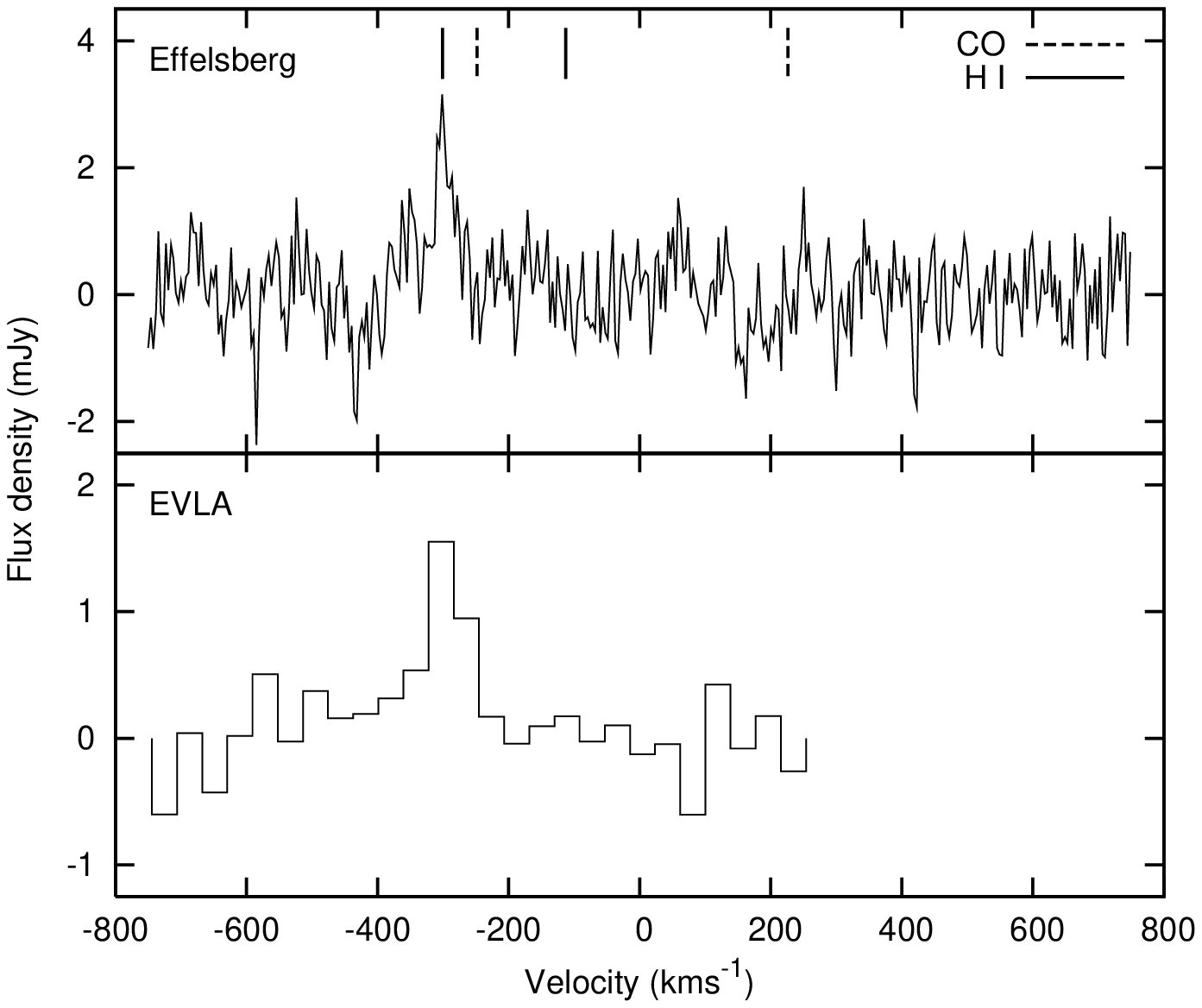}}
\end{picture}
\caption{The 6.1~GHz H$_2$O spectra of the lensed quasar MG~J0414+0534. 
The velocity scale is relative to redshift 2.639 using the optical
velocity definition in the heliocentric frame. The solid lines mark
the H\,{\sc i} absorption components ($-$301\,km\,s$^{-1}$ $\pm$
13\,km\,s$^{-1}$ and $-$113\,km\,s$^{-1}$ $\pm$ 13~km\,s$^{-1}$) and
the dashed lines indicate the peaks of CO emission
($-$238\,km\,s$^{-1}$ $\pm$ 70\,km\,s$^{-1}$ and $+$226\,km\,s$^{-1}$
$\pm$ 70~km\,s$^{-1}$).  Top panel: The combined spectrum taken with
the Effelsberg 100~m radio telescope on 16 July and 14 September
2007. The total on-source integration time was 14~h. The spectra were
formed with a 1024~channel autocorrelator, which provided a channel
width of 3.83~km\,s$^{-1}$.  The rms noise level of the spectrum is
0.6~mJy~channel$^{-1}$.  Bottom panel: The spectrum of lensed images
A1 and A2 of MG~J0414+0534 taken with the EVLA using nine of the
upgraded 25~m antennas during 24, 28, and 30 September and 1 and 7
October 2007. The usable observing time was 12~hours on-source. The
spectrum has 32 channels with a spectral resolution of
38.4~km\,s$^{-1}$~channel$^{-1}$. The rms noise level is
0.3~mJy~channel$^{-1}$.}
\label{spec}
\end{center}
\end{figure}

\begin{figure}
\begin{center}
\setlength{\unitlength}{1cm}
\begin{picture}(6,7)
\put(-5,-8){\includegraphics{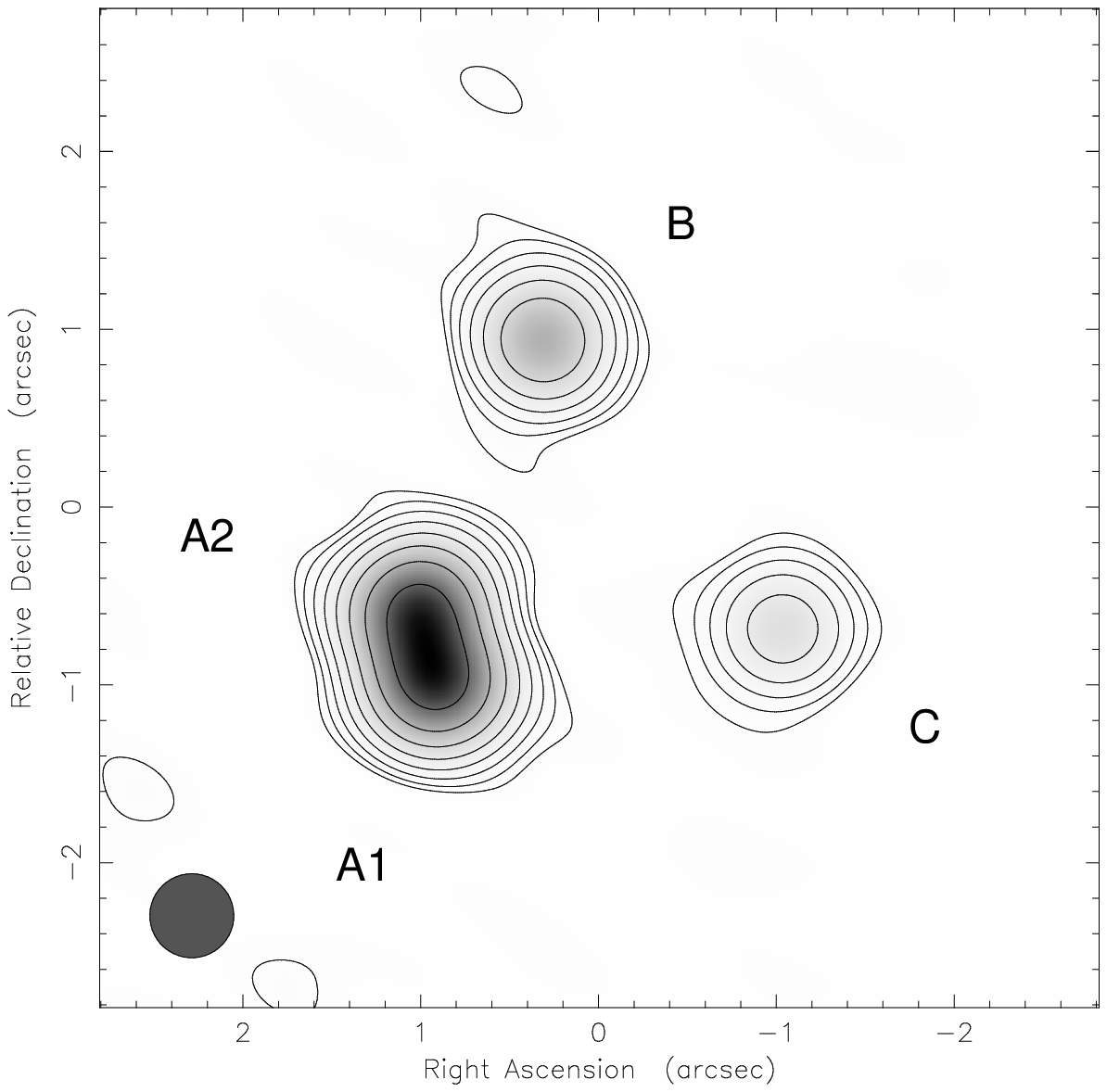}}
\end{picture}
\caption{The 6.1~GHz EVLA radio continuum image of MG~J0414+0534. The water maser emission line shown in Figure 1 was obtained by integrating over lensed images A1 and A2. The data were taken with the EVLA in BnA configuration during
September and October 2007. The observations (25\,MHz bandwidth, 32
spectral channels) were amplitude calibrated relative to 3C\,48, for
which we adopted a flux density of 4.35\,Jy, and were bandpass
corrected by observing the calibrator PKS J0423--0120 approximately every
30 minutes. Natural weighting was used to increase the 
 sensitivity
to the weak line emission during imaging. The spectral line data were
averaged over the inner 24 spectral channels to form a continuum
dataset, which was self-calibrated and then deconvolved using
CLEAN. The resulting antenna phase solutions were applied to the
spectral line data resulting in a line dataset which was phase
referenced to the continuum. For simplicity, and because the source is
unresolved in the E-W direction, the continuum image was restored
using a circular Gaussian beam with a full width at half-maximum (FWHM) of 0.47~arcsec. Note that
the synthesised beam of the EVLA has a FWHM of 0.93 arcsec $\times$
0.47 arcsec at a position angle of 79 degrees east of north. The total radio continuum flux density of the four lensed images is
0.56~Jy $\pm$ 0.06~Jy, which is in good agreement with previous
measurements$^{16}$. The image contours are ($-$3, 3, 6, 12, 24, 48,
96, 192, 384) $\times$ 0.43~mJy~beam$^{-1}$. The rms noise is
0.43\,mJy~beam$^{-1}$. }
\label{map}
\end{center}
\end{figure}

\end{document}